\documentstyle{mn}

\newif\ifAMStwofonts

\ifoldfss
  \ifCUPmtlplainloaded \else
    \NewTextAlphabet{textbfit} {cmbxti10} {}
    \NewTextAlphabet{textbfss} {cmssbx10} {}
    \NewMathAlphabet{mathbfit} {cmbxti10} {} 
    \NewMathAlphabet{mathbfss} {cmssbx10} {} 
  \fi
  \ifAMStwofonts
    \ifCUPmtlplainloaded \else
      \NewSymbolFont{upmath} {eurm10}
      \NewSymbolFont{AMSa} {msam10}
      \NewMathSymbol{\upi}     {0}{upmath}{19}
      \NewMathSymbol{\umu}     {0}{upmath}{16}
      \NewMathSymbol{\upartial}{0}{upmath}{40}
      \NewMathSymbol{\leqslant}{3}{AMSa}{36}
      \NewMathSymbol{\geqslant}{3}{AMSa}{3E}

       \let\le=\leqslant
       
    \fi
  \fi
\fi 

\ifnfssone
  \newmathalphabet{\mathit}
  \addtoversion{normal}{\mathit}{cmr}{m}{it}
  \addtoversion{bold}{\mathit}{cmr}{bx}{it}
  \newmathalphabet{\mathbfit} 
  \addtoversion{normal}{\mathbfit}{cmr}{bx}{it}
  \addtoversion{bold}{\mathbfit}{cmr}{bx}{it}
  \newmathalphabet{\mathbfss} 
  \addtoversion{normal}{\mathbfss}{cmss}{bx}{n}
  \addtoversion{bold}{\mathbfss}{cmss}{bx}{n}
  \ifAMStwofonts
    \ifCUPmtlplainloaded \else
      \UseAMStwoboldmath
      \makeatletter
      \new@mathgroup\upmath@group
      \define@mathgroup\mv@normal\upmath@group{eur}{m}{n}
      \define@mathgroup\mv@bold\upmath@group{eur}{b}{n}
      \edef\UPM{\hexnumber\upmath@group}
      \new@mathgroup\amsa@group
      \define@mathgroup\mv@normal\amsa@group{msa}{m}{n}
      \define@mathgroup\mv@bold\amsa@group{msa}{m}{n}
      \edef\AMSa{\hexnumber\amsa@group}
      \makeatother
      \mathchardef\upi="0\UPM19
      \mathchardef\umu="0\UPM16
      \mathchardef\upartial="0\UPM40
      \mathchardef\leqslant="3\AMSa36
      \mathchardef\geqslant="3\AMSa3E

       \let\le=\leqslant

    \fi
  \fi
\fi 

\ifnfsstwo
  \DeclareMathAlphabet{\mathbfit}{OT1}{cmr}{bx}{it}
  \SetMathAlphabet\mathbfit{bold}{OT1}{cmr}{bx}{it}
  \DeclareMathAlphabet{\mathbfss}{OT1}{cmss}{bx}{n}
  \SetMathAlphabet\mathbfss{bold}{OT1}{cmss}{bx}{n}
  \ifAMStwofonts
    \ifCUPmtlplainloaded \else
      \DeclareSymbolFont{UPM}{U}{eur}{m}{n}
      \SetSymbolFont{UPM}{bold}{U}{eur}{b}{n}
      \DeclareSymbolFont{AMSa}{U}{msa}{m}{n}
      \DeclareMathSymbol{\upi}{0}{UPM}{"19}
      \DeclareMathSymbol{\umu}{0}{UPM}{"16}
      \DeclareMathSymbol{\upartial}{0}{UPM}{"40}
      \DeclareMathSymbol{\leqslant}{3}{AMSa}{"36}
      \DeclareMathSymbol{\geqslant}{3}{AMSa}{"3E}

       \let\le=\leqslant

    \fi
  \fi
\fi 

\ifCUPmtlplainloaded \else
  \ifAMStwofonts \else 
    \def\upi{\pi}
    \def\umu{\mu}
    \def\upartial{\partial}
  \fi
\fi

\title{The Canada-France Redshift Survey X: The Quasar Sample}

\title[CFRS X: The Quasar Sample]
  {Canada-France Redshift Survey X: The Quasar Sample}
\author[David Schade et al.]
 {David Schade,$^{1,4}$
  David Crampton$^{2,4}$
  F. Hammer$^{3,4}$
  O. Le F\`evre$^{3,4}$
  S.J. Lilly$^{1,4}$\\
$^1$Department of Astronomy, University of Toronto, 60 St. George St., 
Toronto, Canada M5S 1A7\\
$^2$Dominion Astrophysical Observatory, National Research Council of Canada,
       Victoria, Canada\\
$^3$DAEC, Observatoire de Paris-Meudon, 92195 Meudon, France\\
$^4$Visiting Astronomer at the Canada-France-Hawaii Telescope which is operated\\
by the National Research Council of Canada, the Centre National de la Recherche\\
Scientifique de France, and the University of Hawaii.\\
}

\date{ }
\pagerange{\pageref{firstpage}--\pageref{lastpage}}
\pubyear{1995}

\def\LaTeX{L\kern-.36em\raise.3ex\hbox{a}\kern-.15em
    T\kern-.1667em\lower.7ex\hbox{E}\kern-.125emX}

\begin{document}

\label{firstpage}

\maketitle

\begin{abstract}

Six objects with broad emission lines and redshifts from
0.48 to 2.07 were discovered among 736 extragalactic objects
in the Canada-France Redshift  Survey  (CFRS).
Although the luminosities
of half of the objects are such that they are in the Seyfert regime
($M_B> -23)$,  all would be designated as quasars in
traditional surveys. Since the only selection 
criterion was that $17.5\le I_{AB} \le 22.5$, or approximately $B < 23$
(assuming a continuum power-law slope $\alpha=-0.5$), these quasars
represent an unbiased, flux-limited sample. Although uncertain, 
the implied surface density, 200$^{+120}_{-80}$ deg$^{-2}$ is the highest yet
measured, and is in good agreement with extrapolations from other faint surveys and  the evolving luminosity function models of Boyle (1991).
The distributions of the continuum properties, emission-line
strengths, etc., of the quasars do not differ significantly from
those of quasars selected by other means, and therefore they would have
been detected in most traditional surveys.  Three of the quasars
may be associated with clusters or large structures of galaxies at
z $>$ 1. 

\end{abstract}
 

\section{Introduction}

 The Canada-France Redshift Survey (CFRS) is 
 aimed at the detection of faint galaxies among objects
with 17.5$\le I_{AB} \le$ 22.5, with no
discrimination  against any other properties.
The final catalog comprises 943 objects, of which 736 are
extragalactic.  
The most common methods used to
produce samples of quasars or active galactic nuclei (AGN)
exploit their unusual colours, strong emission lines, variability, radio
emission, X-ray emission, etc., and therefore are always 
biassed in some sense. Hewett and Foltz (1994) stress
that if a completely unbiassed sample is not possible, an accurate
assessment of the probability of detection as a function of absolute
magnitude, redshift and spectral energy distribution must be
made, and this is frequently lacking.
Even though very effective and efficient survey techniques have been
developed, there is always a danger that there
may exist a subset of the population
that might have been overlooked. Indeed, Webster(1995)
has recently suggested that many more red quasars exist
than previously supposed. Although the results of
surveys based on  techniques other than the popular
UVX method can be examined to see if any
such red population exists, surveys like CFRS 
can be considered
as the ultimate check since spectroscopic
observations are made for every object regardless of colour, morphology
or spectroscopic properties. For example, the CFRS survey
could detect  red, radio-quiet, X-ray quiet 
quasars should they exist. On the other
hand, since quasars constitute only $\sim$ 1\% of the surface
density of objects on the sky at these magnitudes, CFRS
cannot be expected to be an efficient survey for quasars.

The fundamental limit to CFRS as a quasar survey is the signal-to-noise
at which a broad emission line object could be identified.
In this paper, the properties of the six objects which would
commonly be classified as quasars in major surveys are compared with
those from other samples. Although distinction is often made
between quasars and other AGNs, particularly Seyfert galaxies, such
a distinction is not usually made in the basic surveys: any point-like
object with strong, broad emission lines is usually classified as a quasar.
Subsequently, further subdivisions are made based on luminosities and
whether or not the object is extended under typical
ground-based seeing conditions. In this paper, we adopt the
traditional survey definition, and simply refer to
all six strong emission line objects as quasars, partly to avoid confusion
with other galaxies in our survey which exhibit ``AGN activity'' in
their nuclei (e.g., Tresse et al. 1994).

Our CFRS quasar sample, although very small, is relatively unique and hence
it is of interest to compare
it with samples derived by other methods.   Colless et al. (1991)  reported
a similar analysis based on detection of two quasars in a 
faint blue-selected sample, $21 < b_{J} < 23.5$.
 Throughout this paper
it is assumed that
$H_\circ=50$ km sec$^{-1}$ Mpc$^{-1}$ and $q_{\circ}=0.5$.

\section{Observations}

 As part of the Canada-France Redshift Survey, spectra were
obtained of over  1000 objects with 17.5$\le I_{AB} \le$ 22.5 in five
different high galactic latitude fields selected to produce a fair sample
of extragalactic objects (Lilly et al 1995a).
The CFHT MOS spectrograph was used to cover
the spectral region 4200 -- 8600\AA\ with a resolution
of 7\AA\ per CCD pixel, and slits were used which yielded a spectral resolution of
FWHM $\sim$35\AA. $BVIK$ photometry was also obtained for most objects
in the spectroscopic sample (Lilly et al. 1995a). 
 Details of all the observations, 
reduction techniques and sensitivity are presented by Le F\`evre et al. (1995a), Hammer et al. (1995), 
Lilly et al. (1995b) and Crampton et al. (1995). 
Eight one hour exposures were usually obtained of all targets so that
spurious features such as cosmic rays were easily rejected,
and features such as strong emission lines were readily apparent in all the individual spectra.
The final CFRS catalog includes 591 galaxies with z $<$ 1.3, 200 stars, 146
unidentified objects and 6 quasars. Colour and morphological information
indicates that most of the unidentified objects are likely to be
galaxies (Crampton et al. 1995) but 7 objects are
quite compact and have colours not very different from the quasars. 
The minimum equivalent width of emission lines that
 would have been detected in the spectra of these seven objects is $\sim$30\AA.
Emission lines in the observed frame of at most a few percent of quasars
at z $>$ 2 would have been missed, if the equivalent width distributions
of Francis et al. (1992) and Hartwick and Schade (1990) are assumed. 
The Mg II equivalent width distribution indicates that
 only $\sim$10\%  of low redshift (z $<$ 1) quasars
 might have been missed, so it is unlikely that any quasars
reside among the unidentified objects. 
Thus, only 6 quasars were detected among
 a total of 736 extragalactic objects.
The total
 effective area surveyed was 112 arcmin$^2$.

 Table 1 summarizes the properties of the quasars. The first column gives the CFRS number (the first two digits represent the
right ascension of the field). The second and third columns give 2000 
coordinates, followed by the isophotal
I$_{AB}$ magnitude ($I_{AB} = I$ + 0.48), redshift, absolute
magnitude, $(V-I)_{AB}$ colour, spectral index and rest-frame
equivalent widths of strong emission lines. The errors of the
equivalent widths are estimated to be $\pm10$\AA\ for most of the lines,
and $\pm5\AA$ for the lines of CFRS14.0198.
 Absolute magnitudes in the rest-frame B band 
were computed assuming a power-law continuum
slope, $f_\nu=\nu^\alpha$, with $\alpha$ determined for each object
from its $(V-I)_{AB}$ colour corrected for emission line contamination.
 As indicated above, not all  of these ``quasars'' lie above
the canonical division
between quasars and Seyfert galaxies at $M_B = -23$ for ($H_\circ=50$
km sec$^{-1}$ Mpc$^{-1}$). It is interesting to note that
no very high redshift quasars were discovered. Since our targets were
selected by their $I$ magnitudes, there is certainly no selection bias
against high redshift quasars, and since Ly$\alpha$ would not disappear
from our spectroscopic bandpass until z $>$ 6, identification of high redshift
quasars would have been straightforward had they been present.
Spectra
of the six quasars are presented in Figure 1 with the strongest emission lines
marked. 
 Finding charts from our deep $I$ band images are given in Figure 2. At least
three of the quasars may be in groups or large structures of galaxies,
as noted below.

\subsection{Notes on individual quasars}

\noindent
{\bf CFRS00.0207  z=1.352} is surrounded by many faint galaxies of similar magnitude; the five CFRS galaxies indicated in Figure 2 have isophotal
$I_{AB} \sim23.4\pm0.3$mag. It is conceivable that this quasar is embedded in a 
cluster similar to those recently proposed by Hutchings et al. (1994)
and Matthews et al. (1994) around other z $>$ 1 quasars.

\noindent
{\bf CFRS03.0106 z=2.070}  is 2\farcs3 from a  $\sim$1.3 mag fainter galaxy.

\noindent
{\bf CFRS03.0603 z=1.048} has a companion (615) only 16$^{\prime\prime}$ away which has
identical redshift (1.048). In the larger CFRS field in this direction,
there are three more galaxies with redshifts within $\Delta$z = 0.01 
bringing the total to
five. We have spectra of only one other object shown in the 
field; CFRS03.0602 is a star. The image of CFRS03.0603 appears to be slightly 
extended compared to the nearby stars, and the 
measured ``compactness parameter'', Q , (Le Fevre, et al. 1986)
 also indicates that it is resolved. Typically, CFRS point source objects
 have Q $\le$1.3 (Crampton et al. 1995), while Q = 1.55 for this quasar. 
 Since the absolute magnitude 
of this quasar is M$= -23.0$, it is at the border of the canonical Seyfert
-- quasar classification. Nevertheless, since z $\sim$1, it must reside
in a bright host galaxy.

\noindent
{\bf CFRS14.0198 z=1.6034:} The bright object (163) is an M star.

\noindent
{\bf CFRS14.1303 z=0.9859}  appears to be part of a very large structure 
(spread over our entire field corresponding to 6.5 $h_{50}^{-1}$ Mpc
projected dimension for $q_\circ=0$,
and greater than 900 km/s in redshift space)
containing an estimated 30 bright galaxies at
z = 0.985 (Le F\`evre et al. 1994). The nearest of these is CFRS14.1262, only
17$^{\prime\prime}$ away. CFRS14.1275 is at z = 0.763 and 
CFRS 14.1327 is at z = 0.932. Strong lines of Ne III and Ne V are 
present in the spectrum of CFRS14.1303 (Fig. 1).

\noindent
{\bf CFRS14.1567 z=0.4787} is slightly extended, indicating that the host 
galaxy is visible.
Since M$= -22.0$, this quasar is more properly classified as a Seyfert.
Only two of the nearby galaxies have redshifts measured; both
1525 and 1541 have z $\sim$0.74. 

Even though there are indications that at least three of the quasars 
may be located in groups of galaxies, a more rigorous analysis
(Le F\`evre et al 1995b) shows that only the group around 14.1303 is
statistically significant.

\section{Discussion}

\subsection{Surface density}

 The effective area of the CFRS survey is 0.031 deg$^{2}$ so that
the observed 6 quasars  translate to a surface density of 
200$^{+120}_{-80}$ 
 deg$^{-2}$;
the highest surface density observed to date. 
The errors enclose a 1$\sigma$ confidence interval (Gehrels 1986).
The survey magnitude
limit of $I_{AB}\le$22.5 corresponds to $B=23$ for quasars with a power-law
spectral energy distribution ($f_\nu\propto \nu^{\alpha}$) with $\alpha=-0.5$.

 For comparison, the previously-faintest quasar surveys 
reached a limiting magnitude of $B\sim$22.  The Koo and
Kron (1988) survey has a limiting magnitude of m$_{J}$=22.5 ( $B \sim$m$_{J}+0.1$), but spectroscopic followup has not yet
been completed (Majewski, private communication).
 Zitelli et al. (1992, hereafter Z92)
have surveyed 0.35 deg$^{2}$ to m$_{J}=22$ (and a smaller area to m$_{J}$=20.85).
Their survey yields a quasar surface density at m$_{J}$=22 of
 115$\pm$17  deg$^{-2}$ and, for z$<$ 2.2,  a surface density
 of 86$\pm17$ deg$^{-2}$.  Boyle, Jones, and Shanks (1991) surveyed
an area of 0.85 deg$^2$ to a similar limiting magnitude. Although
 their survey suffers from incompleteness for z $>$ 2.2 due
to the UVX selection procedure, the
surface density for quasars with z$<$2.2, $68\pm9$ deg$^{-2}$,
is in reasonable agreement with the Z92 result.
In total,  118 new quasars were discovered in these two surveys. The Z92
result is also consistent with the surface density estimated 
in a review of all major surveys by Hartwick and Schade (1990). 
They derive a surface density of 160 quasars deg$^{-2}$ at $B$ = 22.5 and
 z $<$ 3.3
(or 129 at z $<$ 2.2) by applying a completeness 
correction to the work of Koo and Kron (1988). 
Figure 3 shows a comparison of all of these results for z $<$ 3.3.
The surface density of quasars, log N $<$ B deg$^{-2}$,
from  Hartwick and Schade (1990) (circles) is combined with the
newer results from Boyle et al. (1991) (open triangle) and Z92
 (solid triangle). The faintest point (solid square)
from the  present survey lies on an extrapolation of the data from
previous surveys, although the error
bar is large. 

 An extrapolation of the Z92 quasar surface density using
their slope of 0.40 for the log $N$ versus $M$ relation gives 
an expected number of 300 deg$^{-2}$ 
at $B = 23$ for all redshifts, or  200 deg$^{-2}$ at z $<2.2$.
As noted above, although all of our quasars are at z $<2.2$, we clearly have no bias against
objects at much larger redshifts. Our observed surface density of 200$^{+120}_{-80}$ 
 deg$^{-2}$ is at most 
a mild contradiction to the extrapolation of the Z92
surface density in the sense that  the expected number of low-redshift
(z $<2.2$) objects is observed, but there is some deficit
of high-redshift objects. 

 The faintest X-ray counts where optical identifications
have been made are dominated by active galactic nuclei (Shanks et al 1991).
Analysis of the deepest ROSAT counts yield surface density estimates of
413 X-ray sources deg$^{-2}$ to a limiting flux
of $2.5 \times 10^{-15}$ erg cm$^{-2}$ s$^{-1}$ 
(Hasinger et al. 1993) and fluctuation analysis (Barcons et al. 1994)
yields an estimate of 900 to 1800  discrete sources deg$^{-2}$
brighter than $7 \times 10^{-16}$ erg cm$^{-2}$ s$^{-1}$. These 
counts (in the  0.5-2 KeV) band which might be taken as 
estimates of the quasar surface density are well above our 
own estimate.

\subsubsection{Comparison with Boyle (1991) model}

The expected number of quasars in various redshift bins (e.g,  $0 <$ z $<0.5$, $0.5<$ z $<1.0$, ...) can be computed from the model of Boyle et al. (1991).
We find good agreement between the expectations derived 
from the evolving model and our observations. Over the range $0 < $z$ < 3$ ,
4.98 quasars are predicted by the model compared to 6 observed.
The redshift distribution of the observed quasars also agrees
well with the model prediction, as shown in Figure 4.
Agreement in luminosity is also good: 2.4 objects or $\sim$50\% should be
 AGN with $M_B>-23$, compared to the real sample which has 3, also 50\%. 

 In summary, the redshift and luminosity distributions of the quasars 
in our sample agree very well with the predictions of the model of the
evolving luminosity function of Boyle et al. (1991), demonstrating consistency
of the model with observations in an $M-$z regime which represents a
significant
extrapolation beyond the regime where it was derived.

\subsubsection{Relative numbers of quasars and Seyferts}

 In the preceding, we have made no distinction between luminous
quasars ($M_B<-23$, $H_\circ=50$, $q_\circ=0.5$) and lower luminosity
AGNs or Seyferts. An interesting point made by Z92 is that
the counts of the Seyfert galaxies are rising more steeply (with a slope
of $\sim0.7$ in the log $N$ vs $m$  relation) than the more
luminous objects (whose slope is $\sim 0.4$). They note that the
ratio of Seyfert to quasar counts increases from 0.1 at m$_{J}<20$ to
$\sim 0.3$ at  m$_{J}<21$ and is $\sim 0.5$ at  m$_{J}<22$. In the faintest
half-magnitude bin the ratio reaches unity.
In our very faint sample, the ratio of Seyfert to quasar
counts also equals 1, confirming the trend noted by Z92.

\subsection{Continuum properties}

 The continua of active galactic nuclei (AGN) are frequently characterized
as  power-law $f_\nu \propto \nu^\alpha$ spectra, although this 
is a simplification when the continuum is considered over a large range in wavelength (e.g., Neugebauer et al. 1979). For example, the continuum
shape in the optical-ultraviolet region is dominated by the 
``blue-bump'' feature which  may be the signature of
a hot accretion disk (Elvis et al. 1994). Since many of the
quasars previously studied have been selected because they have blue continua
(e.g., in ultraviolet-excess surveys), there may be
 biases in the statistics related to continuum shape. 
Hence, it is of interest to
examine the continuum slopes in our sample where no such bias
can exist.

 The optical-UV region of quasar spectra are crowded with emission
lines, Balmer continuum emission, and a number of FeII emssion features
(e.g., Francis et al. 1991, hereafter F91). This makes the definition of the
continuum difficult, particularly when only a restricted wavelength range
(and one which varies with redshift) is available. 
In general, our spectra were well-fitted by power laws, but estimates of $\alpha$
made in this way are subject to considerable error, due to errors in the 
centering of any individual object among the $\sim$80 objects 
in our multi-slit masks.
Consequently, the continuum spectral index was estimated from our $V-I$ 
colours (corrected for the contribution of emission lines), assuming a power law. 
The slopes, listed in Table 1, range from $\alpha=0.2$ to 
$\alpha=-2.0$ with a mean of $-1.0$.
The error is estimated to be $\sim \pm 0.2$ in $\alpha$

One of the largest and most complete surveys
with which to compare our results is the Large Bright Quasar Survey (LBQS).
F91 give the distribution 
of power-law slopes measured in the wavelength range
1450 -- 5050\AA\  for 688 LBQS (m $\le18.85$) quasars. They find a mean 
$\alpha = - 0.32$. Francis et al. (1992) found a higher value using
a principal component analysis method 
on a smaller subsample of the LBQS, but since the rest-frame continuum window used was very blue, the  earlier analysis is better matched to our data.
Interestingly, the results from our small sample span nearly the entire
range in continuum slope evident in the much larger sample
of F91. Half of our spectral indices are redder than
$-1$ whereas at most 10\% of those in F91 
are as red as this. Furthermore, 5/6 of our
spectral indices are redder than the median of the
sample of F91. However, the 95\% confidence interval (Gehrels 1986)
for the ratio $R$ of red-to-total objects derived from
our observations is $0.36 < R < 0.996$, 
 so that this result is not highly significant. 
 There is no reason to believe that the survey
technique employed in the LBQS
would have missed the reddest (or any) quasars seen here. On the contrary,
Francis et al. (1992) claim that any quasar with a continuum
bluer than $\alpha\sim -2$ and/or emission lines would have
been detected. Thus, although there is some indication that our
sample is rather red, it does not constitute a population
that would have escaped detection in the LBQS or other surveys.
It should be remembered, however, 
that there are substantial
differences in the luminosity and redshift distributions between our
sample and the LBQS.

\subsection{Emission-line strengths}

The rest-frame
equivalent widths of the three principal emission lines common
to most of our spectra are listed Table 1. The equivalent widths
are consistent with the means given
by F91 and those compiled by Hartwick
and Schade (1990), indicating once again that the quasars in our
sample are not unusual. Since there has always been a lingering concern that
the average emission line strengths of quasars have been biased by their
detection methods, this is reassuring.

\section{Summary}

 The main goal of this paper was to compare the properties of
this very faint (albeit small) sample of quasars
 with the properties of previously known samples. This
sample was selected without any of the biases inherent in
the most common survey techniques and therefore could
 potentially yield new insights into the population.

 The surface density of AGN at $B=23$ is $200^{+120}_{-80}$
 deg$^{-2}$, the highest observed to date.
 This is consistent with an extrapolation of
the faint counts due to Z92 and
Boyle et al. (1991), and also with the lower limit found
by Colless et al. (1991) based on a similar method to ours.
 We find that the space density of these quasars
as a function of redshift and luminosity agrees
very well with the predictions of the
successful evolutionary model of
Boyle (1991), even though our sample is deeper than any of the
samples from which the model is derived.

 The colours of these faint AGN are within the range defined
by previously known samples and would not have escaped
detection in major surveys. We find an apparent surplus of red
objects compared with other studies, but the numbers are small
and the significance is low. The strengths
of the emission lines are entirely consistent with those
found in other surveys. 
 In summary, we find no remarkable differences between any of the
properties of these faint objects and the distributions
of the same properties derived from previous
surveys. All our quasars would have been found by previous
 search techniques.

One of the quasars is in a structure consisting of at
least 30 galaxies at z $\sim0.985$, and there is evidence
that two others may be associated with groups
 of galaxies at a similar redshift.

\newpage

\begin{figure}
\caption{Flux calibrated spectra of the six quasars discovered in the CFRS
survey. The strongest emission features are identified. 
Heavy horizontal lines indicate photometric measurements from
the direct imaging in $V$ and $I$.
Unfortunately, 
the [O II] 3727\AA\ line in the spectrum of CFRS03.0603 is coincident 
with the atmosperic A band, and of CFRS14.1567 with the strong [O I] 5577\AA\
night sky emission line. Consequently their intensities are unreliable.}
\end{figure}

\begin{figure}
\caption{Finding charts from deep $I$ band images of the six quasars. The images are 31$^{\prime\prime}$ on a side, with N to the top, E to the left, 
and the quasar is in the center of each field. Other
catalogued objects in the fields are identified by the last 
digits of their CFRS catalog numbers. }
\end{figure}

\begin{figure}
\caption {The cumulative surface density of quasars projected on the sky. The
data represented by open circles are from the compilation of
Hartwick and Schade (1990), the open triangle is from Boyle et al. (1991)
the solid triangle from  Z92. The faintest point (solid square)
is from the  present survey. }
\end{figure}

\begin{figure}
\caption {The redshift histogram of the CFRS quasar sample (hatched area)
compared to the prediction from the evolving luminosity function model
of Boyle (1991).}
\end{figure}

\label{lastpage}

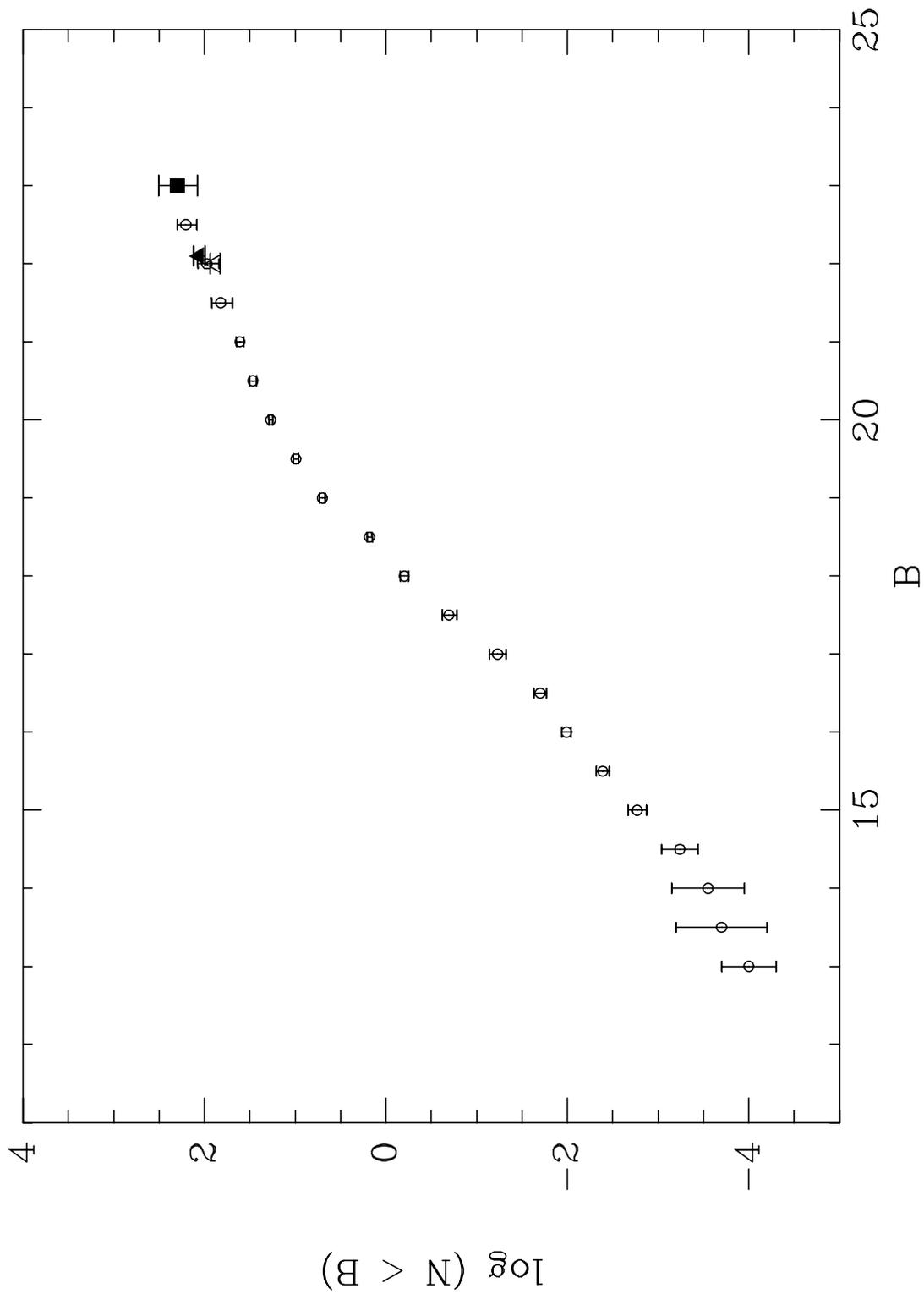
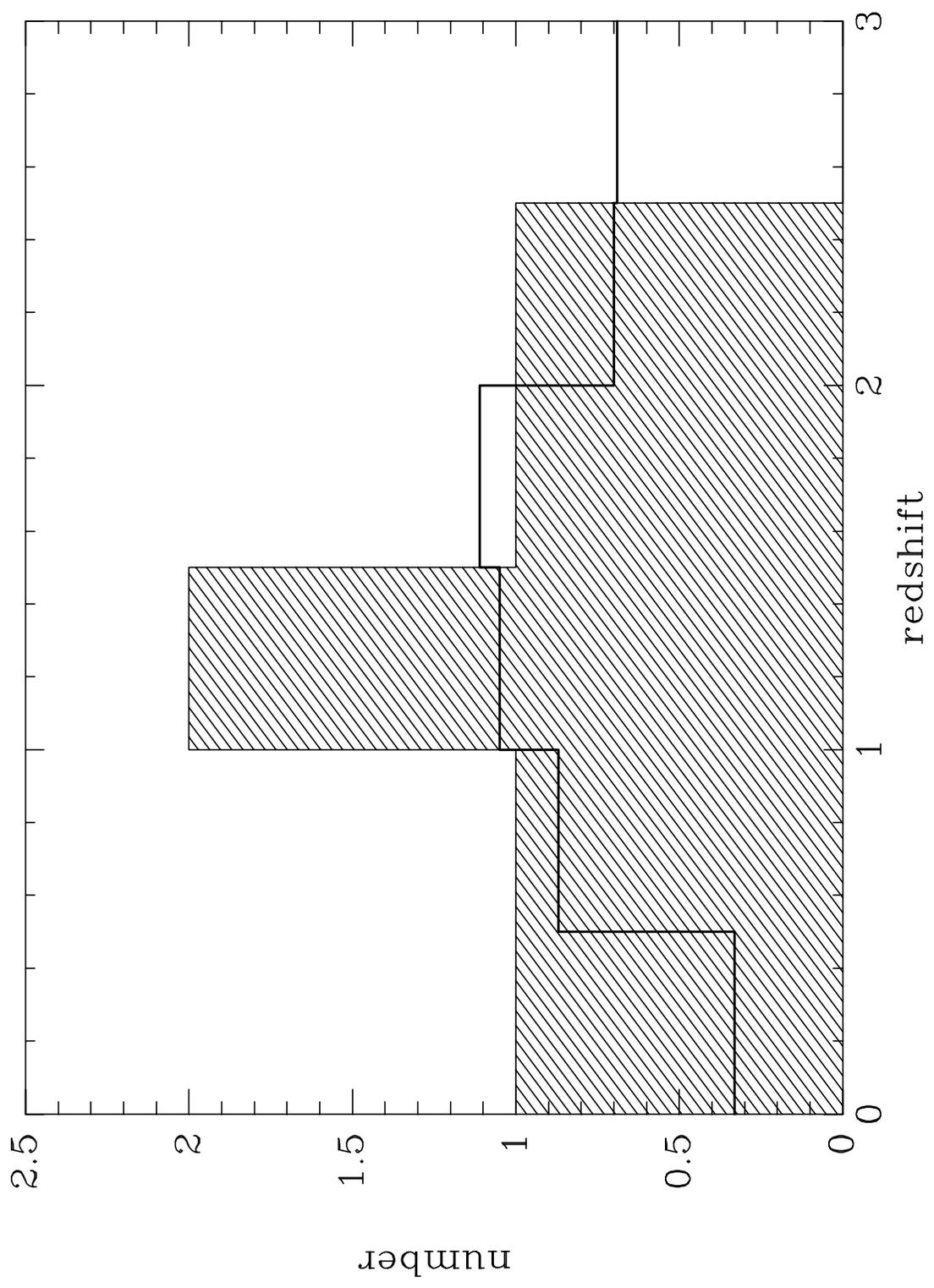
\end{document}